\documentclass{aa}
\usepackage{bbding}
\usepackage{float}

\usepackage{graphicx}

\usepackage{amsmath}
\usepackage{amsfonts} 
\usepackage[all]{xy}
\usepackage{txfonts}

\usepackage{txfonts}
\begin{document}
 \title{Evolution of Relative Magnetic Helicity: New Boundary Conditions for the Vector Potential}
   \author{
Shangbin Yang \inst{1,2,3}
\and J\"org B\"uchner \inst{2}
\and Jan Sk\'ala \inst{2,4}
\and Hongqi Zhang   \inst{1} }

   \institute{Key Laboratory of Solar Activity, National
Astronomical Observatories, Chinese Academy of Sciences, 100012
Beijing, China \\
             \email{yangshb@bao.ac.cn}
         \and
     Max-Planck Institute for Solar System Research, 37077 G\"ottingen, Germany
     \and
     University of Chinese Academy of Sciences, 100049 Beijing,China
     \and
     Astronomical Institute of Czech Academy of Sciences, 25165 Ondrejov, Czech Republic
     \and
     University J. E. Purkinje, 40096 Usti nad Labem, Czech Republic
}

\abstract
{For a better understanding of the dynamics of the solar corona it is important to analyze the evolution of the helicity of the magnetic field. Since the helicity cannot be directly determined by observations  we recently proposed a method to calculate the relative magnetic helicity in a finite volume for a given magnetic field in \citep{Yang13}, which however required the flux to be balanced separately on all the sides of the considered volume.
}
{Development of a scheme to obtain the vector potential in a volume without the above restriction at the boundary. Study of the dissipation and escape of relative magnetic helicity from an active region.
}
{In order to allow finite magnetic fluxes through the boundaries, a Coulomb gauge is constructed that allows for global magnetic flux balance. The property of sinusoidal function is used to obtain the vector potentials at the twelve edges of the considered rectangular volume extending above an active region. We tested and verified our method in a theoretical fore-free magnetic field model.
}
{We apply the new method to the former calculation data and found a difference of less than 1.2\%. We also applied our method to the magnetic field above active region NOAA 11429 obtained by a new photospheric-data-driven MHD model code GOEMHD3.  We analyzed the magnetic helicity evolution in the solar corona using our new method. It was found that the normalized magnetic helicity ($H_{\rm}/\Phi^2$) is equal to -0.038 when fast magnetic reconnection is triggered.  This value is comparable to the previous value (-0.029) in the MHD simulations when magnetic reconnection happened and the observed normalized magnetic helicity (-0.036) from the eruption of newly emerging active regions. We found that only 8\% of the accumulated magnetic helicity is dissipated after it is injected through the bottom boundary. This is in accordance with the Woltjer conjecture. Only 2\% of magnetic helicity injected from the bottom boundary escapes through the corona. This is consistent with the observation of magnetic clouds, which could take away  magnetic helicity into the interplanetary space, in the case considered here, several halo CMEs and two X-class solar flares origin from this active region.
}
{}


\authorrunning{Yang  et al.}
\titlerunning{Evolution of RMH: New Boundary Conditions for the Vector Potential}

\maketitle

\section{Introduction}
Magnetic helicity is a key geometrical parameter to describe the
structure of solar coronal magnetic fields \citep[
e.g.][]{ber99}. Magnetic helicity in a volume $V$ can be determined as
\begin{equation}
\centering
{
 {H_{\rm M}=\int_{V}\mathbf{A}\cdot\mathbf{B}dV},
 \label{eq:helicityDefine}
}
\end{equation}
where {\bf A} is a vector potential and {\bf B} is  magnetic field
in this volume. Magnetic helicity $H_{\rm M}$ is conserved in an ideal
magneto-plasma \citep{wol58}. As long as the overall magnetic
Reynolds number is large, it is still approximately
conserved, even in the case of relatively slow magnetic
reconnection taking place \citep{Tay74,BER84}. Since the vector potential {\bf A} is not uniquely
defined $H_{\rm M}$ is not gauge-invariant. The magnetic helicity in a volume has a
well-determined value only when the magnetic field at the boundary is exclusively
tangential, {\it i.e.} if ${\bf B} \cdot
\hat{\bf n}|_S = 0$. On the other hand, Berger and Field (1984) have
shown that in the case of boundaries open to magnetic flux
penetration, instead, relative magnetic helicity ($H_{\rm R}$),
given by the Finn-Antonsen (1985) formula

\begin{equation}
H_{\rm R} = \int {\left( {\mathbf A + \mathbf A_{\rm p} } \right)
\cdot \left( {\mathbf B - \mathbf P} \right)} dV ,
\label{eq:RHvolume}
\end{equation}
is gauge-invariant if the magnetic field is chosen as a
reference field such that ${\mathbf{P}\cdot\widehat{\rm n}|_{S}=\mathbf{B}\cdot\widehat{\rm
n}|_{S}}$. It is customary to choose potential field $(\nabla\times\mathbf{P}=0)$ as the reference
field.

The concept of magnetic helicity has successfully been applied to characterize solar coronal process as well as the solar dynamo to interpret solar observations and corona simulations. \cite{Cha01} obtained the relative helicity by applying Fourier transform (FFT) method and local correlation tracking (LCT) to MDI data. Since then the helicity calculation method has been developed further \citep{Dem03,Par05,Sch08} by improving the helicity density map and velocity tracking techniques. The analysis of magnetic helicity based on observations and simulations has also been developed such as in the analysis of magnetic
helicity injection in the course of flux emergence \citep{Yang09a,liu14},the correlation between helicity
change and solar eruption \citep{ZhangY08,Park08,Yang15}, magnetic energy-helicity relation analysis in solar eruptions by \citep{Tzi12}, magnetic helicity distribution in different scales in the solar dynamo \citep{Seehafer03,Pipin14}, magnetic helicity estimation in the interplanetary magnetic cloud
\citep{Dem15}, testing the conjecture of Taylor conjecture in quasi-ideal MHD simulations \citep{par15}. For a recent review of modeling and observations of  magnetic helicity see, {\it e.g.},~\citep{Dem09}.

Although the magnetic helicity is conserved in the fast reconnection process in the close volume, the redistribution of magnetic helicity i.e. helicity transport could still happen, as it is strongly coupled with the magnetic energy release process. For example, the magnetic helicity exchange process has been found between neighboring emerging active regions \citep{Yang09a}. Thus the correlation study between magnetic helicity distribution and evolution and solar eruption become important.
\citet{Zhang06} noted that the accumulation of magnetic helicity in the corona plays a significant role in storing
magnetic energy. They propose that there is an upper bound on the
total magnetic helicity that a force-free field can contain. \cite{Nin05} found that magnetic helicity of CME productive ARs is higher than other ARs. The survey of \citet{Lan07} to helicity accumulation in 393 ARs a revealed that a necessary condition for the occurrence of an X-flare is that the peak helicity flux has a magnitude larger than $3\times10^{36}Mx^2/s$.
In the simulations the debates also exist for the possible upper-bound helicity before the solar eruption happens \citep{Amari04,Jac06}. In these papers, the corresponding normal relative magnetic helicity ($H_{\rm}/\Phi^2$) when the eruption happens reaches approximately -0.16 for case of Kink instability and approximately -0.18 for case torus instability in the simulations of \citep{Fan07}. However, from the observations the range of normalized helicity for the active region is from 0.02 to 0.08 \citep{Lan07,Yang09a,Tian08}. This is  one order of magnitude smaller value than the above simulations. \citet{Yang13} investigate the value of normalized helicity reached 0.0298 just prior to drastic energy release by magnetic reconnection by using the magnetic field above active region NOAA 8210 obtained by a photospheric-data-driven MHD model \citep{Santos11}. \citet{Yang15} studied an emerging and quickly decaying active region (NOAA 9729) in detail as it passed across the solar disk. There was only one CME associated with that active region. This provided a good opportunity to find that the consequences of single CME after the injection of magnetic helicity to the solar corona. The absolute value of normalized magnetic helicity was 0.036 just before solar eruption happened \citep{Yang15}, which is close to the simulation of NOAA 8210 and the theoretical prediction value of \citet{Zhang08} only in a multipolar force-free magnetic field structure. However, whether the relative magnetic helicity  has a upper-bound and how it plays a important role in solar eruptions is still an open question.

Despite of its
important role in the dynamical evolution of solar plasmas, so far
only a few attempts have been made to estimate the helicity of
coronal magnetic fields based on observations and numerical
simulations (see, {\it e.g.}, Thalmann, Inhester, and Wiegelmann,
2011; Rudenko and Myshyakov, 2011). \cite{Yang13} developed a method for calculating the relative magnetic helicity in a finite 3D volume which was applied to a simulated flaring AR 8210 \citep{Santos11}. However, this method required that
the magnetic flux is balanced  on each of the side boundaries
in the considered volume. In this paper, the corresponding scheme is presented which does not require such
restriction on the flux balance. Such method has already been applied in \citet{Val16} and it is the most accurate one among the finite volume methods employing the Coulomb gauge.

 We applied our new method to the magnetic field above active region NOAA 11429 as it was obtained by a photospheric-data-driven MHD model GOEMHD3 \citep{Skala15}.
In Sec.~\ref{sec:old}, we describe
the restriction of vector potential in the previous paper. In
Sec.~\ref{sec:new}, we present the  new scheme to
calculate the vector potentials on the six boundaries. In
sec.~\ref{sec:check}, we use a Non-Linear Force-Free magnetic field model \citep{low90}  and MHD simulation model to check our scheme. The  summary and some discussions are given in
Sec.~\ref{sec:summary}.


\section{Restricted method to obtain ${\bf A}_{\rm p}$ and {\bf A} at the boundaries}
\label{sec:old}
The computation of $H_R$ of Eq.2 requires the knowledge of $\vec{A}$, $\vec{A_p}$, and $\vec{P}$ from the known $\vec{B}$ in the volume $V$. We adopt the Coulomb gauge for the vector potentials, see Section 2.2 of \citet{Yang13} for details.
Let us define a finite three-dimensional (3-D) rectangular volume in
Cartesian coordinates. The magnetic field ${\bf B}(x, y, z)$ is given
in this volume. The volume is restricted to $x = [0, l_x]$, $y =
[0, l_y]$, and $z = [0, l_z]$.

In order to solve for the vector potentials, the boundary conditions must be specified. First, one has to provide the values of ${\bf A}_{\rm p}$ and {\bf A}
on all six boundaries ($x = 0,l_x; y = 0,l_y; z = 0,l_z$).
Taking the bottom boundary ($z = 0$) as an example, we define a new scalar
function $\varphi (x, y)$ that determines  the vector potential
${\bf A}_{\rm p}$ of potential magnetic field {\bf P} corresponding to  ${\bf B}$ on this
boundary as follows:
\begin{equation}
 {
 A_{\rm p\it x}  =  - \frac{{\partial \varphi }}{{\partial y}}, \qquad
A_{\rm p \it y}  = \frac{{\partial \varphi }}{{\partial x}}, \qquad
 \left.{A_{\rm p \it z} } \right|_{z = 0}  = 0.
 \label{eq:sidebou}
 }
\end{equation}
According to the definition of the vector potential, the scalar
function $\varphi (x, y)$ satisfies the Poisson equation:
\begin{equation}
\Delta \varphi(x,y)  = B_z (x,y,z = 0).
 \label{eq:bouPoisson}
\end{equation}
In our previous work \citep{Yang13} we set the values of
$\partial\varphi/\partial n$ at the four edges of the
plane $z=0$ to zero in Equation~(\ref{eq:bouPoisson}).
 As a consequence, the corresponding magnetic flux at the
boundary also vanish according to Amp\`ere's law. The values of
${\bf A}_{\rm p}$ on the other five boundaries can be obtained in
the same way. For the vector potential {\bf A} at all boundaries
the same values are taken as for ${\bf A}_{\rm p}$. When the
magnetic fluxes through the six boundaries are finite, one should
calculate the values of vector potentials at the twelve edges of the
three-dimensional volume to obtain a Neumann boundary condition for
the Poisson equation (\ref{eq:bouPoisson}) at each side boundary.
In next section, we will introduce a scheme to calculate the vector
potentials at the twelve edges.
\section{General method to obtain $\mathbf A_p$ and $\mathbf A$ at the
boundaries} \label{sec:new}
\begin{figure}[htb]
 \includegraphics[angle=0,scale=0.5]{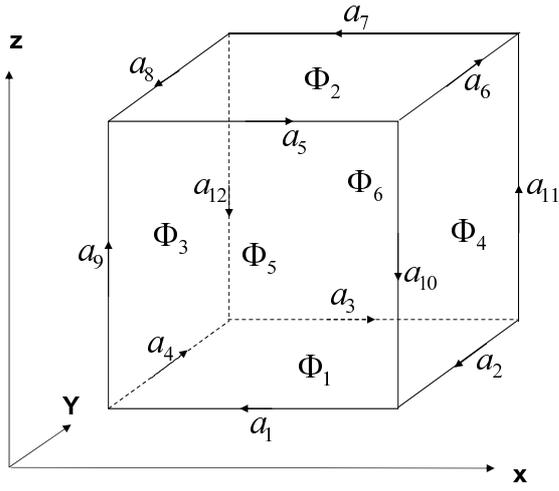}
  \caption{Magnetic flux $\Phi_i~(i=1,...,6)$ through the six boundaries and integration pathes $a_i~(i=1,...,12)$ along the twelve edges of the volume.}
  \label{fig:sketch}
\end{figure}

In order to determine $\mathbf A_p$, we define the magnetic flux
$\Phi_i~(i=1,...,6)$ respectively at each side boundary
($z=0;~z=l_z;~x=0;~x=l_x;~y=0;~y=l_y$). The integrals of $\int{\bf
A}_{\rm p} \cdot \rm d{\bf l}$ at the twelve edges are defined as
$a_i~(i=1,...,12)$. The twelve integrals and the corresponding
directions are depicted in Fig.~\ref{fig:sketch}.
Applying Amp\`ere's law to the six boundaries independently, one obtain a linear system of equations for obtaining the integral value $a_i$ :
\begin{equation}
{\rm{{\bf {\bf T}}{\bf X} = {\bf F}},}
 \label{eq:lineq}
\end{equation}
where $\textrm{\bf F}=(\Phi_1,\Phi_2,\Phi_3,\Phi_4,\Phi_5,\Phi_6)^T$,
$\textrm{\bf X}=(a_1,a_2,a_3,...,a_{12})^T$ and $\rm{\hat{\bf {T}}}$ is a 6$\times$12 matrix given by
\begin{equation}
\tiny
\rm{\hat{\bf {T}}}=
\left[
\begin{array}{cccccccccccc}
1   &  1  & 1  &  1 & 0   & 0   &  0  &  0  &  0  &  0 &  0 &  0 \\
0   &  0  & 0  &  0 & 1   & 1   &  1  &  1  &  0  &  0 &  0 &  0 \\
0   &  0  & 0  &{-1}& 0   & 0   &  0  &{-1} &  1  &  0 &  0 &  1 \\
0   &{-1} & 0  &  0 & 0   &{-1} &  0  &  0  &  0  &  1 &  1 &  0 \\
{-1}&  0  & 0  &  0 &{-1} &  0  &  0  &  0  &{-1} &{-1}&  0 &  0 \\
0   &  0  &{-1}&  0 & 0   &  0  &{-1} &  0  &  0  &  0 &{-1}&{-1}\\
\end{array}
\right]. \label{eq:matrix}
\end{equation}
The six rows in this matrix are not linearly
independent because the magnetic field is divergence free and the
sum of $\Phi_i$ through all six boundaries vanishes. Therefore, there are no unique solutions for the
twelve integrals $a_i$.  We use the freedom in the gauge to construct twelve independent
equations to obtain unique solutions for $a_i$.  We remove the last row of Eq (6) and
add another seven rows to define a
new matrix $\rm{\hat{\bf {T'}}}$ to make sure the determinant of $\rm{\hat{\bf {T'}}}$ is not zero without limiting
the validity of the solution for $a_i$. we choose the following seven rows  to construct the new matrix  $\rm{\hat{\bf {T'}}}$:
\begin{equation}
\tiny
\rm{\hat{\bf {T'}}}=
\left[
\begin{array}{ccccccccccccc}
1   &  1  & 1  &  1 & 0   & 0   &  0  &  0  &  0  &  0 &  0 &  0 \\
0   &  0  & 0  &  0 & 1   & 1   &  1  &  1  &  0  &  0 &  0 &  0 \\
0   &  0  & 0  &{-1}& 0   & 0   &  0  &{-1} &  1  &  0 &  0 &  1 \\
0   &{-1} & 0  &  0 & 0   &{-1} &  0  &  0  &  0  &  1 &  1 &  0 \\
{-1}&  0  & 0  &  0 &{-1} &  0  &  0  &  0  &{-1} &{-1}&  0 &  0 \\
1   &  0  &{-1}&  0 &  0  &  0  &  0  &  0  &  0  &  0 &  0 &  0 \\
0   &  1  & 0  &{-1}&  0  &  0  &  0  &  0  &  0  &  0 &  0 &  0 \\
0   &  0  & 1  &  0 & {-1}&  0  &  0  &  0  &  0  &  0 &  0 &  0 \\
0   &  0  & 0  &  1 &  0  & {-1}&  0  &  0  &  0  &  0 &  0 &  0 \\
0   &  0  & 0  &  0 &  1  &  0  & {-1}&  0  &  0  &  0 &  1 &  0 \\
0   &  0  & 0  &  0 &  0  &  1  &  0  & {-1}&  0  &  0 &  0 &  0 \\
0   &  0  & 0  &  0 &  0  &  0  &  1  &  0  & {-1}&  0 &  0 &  0 \\
\end{array} \right]. \label{eq:matrixn}
\end{equation}
This choice for the lower seven rows of $\rm{\hat{\bf {T'}}}$ is not unique, but any influence on the gauge will be removed later.
One can calculate that  the determinant of $\hat{\rm{\bf T'}}$ does not
vanish. According to Cramer's rule, an unique solution of
the new linear equation
\begin{equation}
{\rm{\hat{\bf T'}}\rm{\bf X} =\rm{\bf F'},}
 \label{eq:lineqn}
\end{equation}
can exists if $\rm{\bf X}=(a_1,a_2,a_3,...,a_{12})^T$ and $\rm{\bf
F'}=(\Phi_1,\Phi_2,\Phi_3,\Phi_4,\Phi_5,0,0,0,0,0,0,0)^T$.
Then we can obtain the integrals of $a_i$ at the twelve edges such that Gauss theorem applied to the Coulomb vector potentials in the volume $V$ is satisfied. Since only the integral values $a_i$ are relevant in Eq.(3), the vector potentials at
the twelve edges can be obtained representing the components of ${\bf A}_{\rm p}$ in the following way:
\begin{equation}
\setlength{\jot}{10pt}
\begin{split}
 {\rm{A}}_{{\rm{px}}} \left( {a_i } \right) = \frac{{\pi a_i }}{{2L_x }}\sin ({{\pi x} \mathord{\left/
 {\vphantom {{\pi x} {L_x }}} \right.
 \kern-\nulldelimiterspace} {L_x }}),i = 1,3,5,7 \\
 {\rm{A}}_{{\rm{py}}} \left( {a_i } \right) = \frac{{\pi a_i }}{{2L_y }}\sin ({{\pi y} \mathord{\left/
 {\vphantom {{\pi y} {L_y }}} \right.
 \kern-\nulldelimiterspace} {L_y }}),i = 2,4,6,8 \\
 {\rm{A}}_{{\rm{pz}}} \left( {a_i } \right) = \frac{{\pi a_i }}{{2L_z }}\sin ({{\pi z} \mathord{\left/
 {\vphantom {{\pi z} {L_z }}} \right.
 \kern-\nulldelimiterspace} {L_z }}),i = 9,10,11,12
 \end{split}
 \label{eq:api}
\end{equation}

Note that such ${\bf A}_{\rm p}$ by construction vanishes
at the ends of every edge i.e. at every corner of the box as required by
Eq.~(\ref{eq:sidebou}).  Now we solve the Poisson equations (\ref{eq:bouPoisson}) to obtain ${\bf A}_{\rm p}$ at the six boundaries. The vector potential {\bf A} at the six boundaries is equal to ${\bf A}_{\rm p}$. The following procedure corresponds to the Sec. 2.2 and Sec 2.3 of \cite{Yang13}.  Using the boundary conditions above, $\mathbf A_p$ and $\mathbf A$ in the volume are obtained by solving the following Poisson and Laplace equations:
\begin{equation}
 {
\Delta \mathbf A_{\rm p} = 0
 }
 \label{eq:poission0}
\end{equation}
everywhere in the volume if ${\bf A}_{\rm p}$ satisfies the Coulomb
gauge. The vector potential {\bf A} of the original magnetic field
{\bf B} satisfies the Poisson equation
\begin{equation}
\Delta \mathbf A=-\mathbf{J}, \label{eq:poission1}
\end{equation}
where ${\bf J} = \mu_0 {\bf j}$ denotes the current density.
In order to remove numerical errors and imperfections in the gauge, we implement a projection method that removes violations of the solenoidal constraint in the volume. In particular, we introduce  a solenoidal
modification vector $\nabla\times {\bf M}$ satisfying the following
condition:
\begin{equation}
\nabla  \times \left( {\nabla  \times \mathbf M} \right) = \mathbf P
- \nabla \times \mathbf A_{\rm p}.
 \label{eq:Mfunction}
\end{equation}
 The components of {\bf M} satisfy the three Poisson equations:
\begin{equation}
\begin{array}{*{20}{c}}
{\left\{ \begin{array}{l}
\Delta {M_z} = {\left( {\nabla  \times {{\vec A}_p}} \right)_z} - {P_z}\\
{M_z}\left( {z = 0,{l_z}} \right) = 0\\
\frac{{\partial {M_z}}}{{\partial x}}\left( {x = 0,{l_x}} \right) = 0\\
\frac{{\partial {M_z}}}{{\partial y}}\left( {y = 0,{l_y}} \right) = 0
\end{array} \right.,}&{\left\{ \begin{array}{l}
\Delta {M_y} = {\left( {\nabla  \times {{\vec A}_p}} \right)_y} - {P_y}\\
{M_y}\left( {y = 0,{l_y}} \right) = 0\\
\frac{{\partial {M_y}}}{{\partial x}}\left( {x = 0,{l_x}} \right) = 0\\
\frac{{\partial {M_y}}}{{\partial z}}\left( {z = 0,{l_z}} \right) = 0
\end{array} \right.,}\\
{\left\{ \begin{array}{l}
\Delta {M_x} = {\left( {\nabla  \times {{\vec A}_p}} \right)_x} - {P_x}\\
{M_x}\left( {x = 0,{l_x}} \right) = 0\\
\frac{{\partial {M_x}}}{{\partial y}}\left( {y = 0,{l_y}} \right) = 0\\
\frac{{\partial {M_x}}}{{\partial z}}\left( {z = 0,{l_z}} \right) = 0
\end{array} \right..}&{}
\end{array}
\label{eq:Mzeqs}
\end{equation}
Finally, in order to remove the residual errors in the solenoidal property of ${\bf A}_{\rm
p}$, we introduce  scalar
field $\phi (x, y, z)$ which satisfies the following Poisson
equation:
\begin{equation}
\left\{ {\begin{array}{*{20}c}
   {\Delta \phi  =  - \nabla  \cdot \mathbf A_{\rm p} }  \\
   {\left. {\frac{{\partial \phi }}{{\partial {n}}}} \right|_s  = \left. { - \left( {\nabla  \times \mathbf M} \right) \cdot \mathord{\buildrel{\lower3pt\hbox{$\scriptscriptstyle\frown$}}
\over {\rm n}} } \right|_s }  \\
\end{array}} \right..
 \label{eq:phi}
\end{equation}
After solving Eq.~\ref{eq:phi}, one obtains a new modified vector potential ${\bf A}_{\rm
p}'$ is represented as
\begin{equation}
\mathbf A'_{\rm p}  = \mathbf A_{\rm p}  +\nabla \times \mathbf M +
\nabla \phi. \label{eq:modAp}
\end{equation}
and inn the same way, one can obtain the corrected vector potential ${\bf
A}'$ just by replacing the right-hand-side term of Eq.~(\ref{eq:Mzeqs}) as
$\nabla\times {\bf A} - {\bf B}$. Now one can calculate the relative helicity in the
volume according to Eq.~(\ref{eq:helicityDefine}). As already mentioned when introducing $\rm{\hat{\bf {T'}}}$, the choice of the lower seven rows of $\rm{\hat{\bf {T'}}}$ and the final seven elements of $F'$ is not unique. 
\section{Testing the scheme}
\label{sec:check} For testing the new scheme to obtain the vector
potentials at the boundaries, we use analytical nonlinear
force-free fields of \citet{low90}. We utilized the model labeled
$P_{1,1}$ with $l = 0.3$ and $\Phi = \pi/2$ in the notation of their
paper. We calculated the magnetic field on a uniform grid of
$64\times 64\times 64$.

We first calculate the magnetic fluxes $\Phi_0$ through
the six boundaries and substitute it to equation~(\ref{eq:lineqn}) to obtain the integrals
$a_i$ along the twelve edges of the 3D volume. Then we substitute $a_i$
into equation~(\ref{eq:api}) respectively to obtain the boundary value for
solving the Poisson equation~(\ref{eq:bouPoisson}) at the
six boundaries. After we obtained ${\bf A}_{\rm p}$ on the six
boundaries, we calculate the magnetic fluxes
 $\Phi$ from the computed vector potential using
 the relation between the vector potential and the
magnetic field: ${\bf B}\cdot \hat{\rm n}=\nabla \times {\bf A}_{\rm
p} \cdot \hat{\rm n}$. Table.~\ref{table:check} represents the
quantitative result after applying the above scheme. As one can see in the table the calculated magnetic fluxes through the six boundaries by using our new scheme is actually better flux-balanced than the theoretical model. Such small errors are partly due to the fact that the total magnetic flux of the
analytical model does not completely vanish. It is required that the total magnetic flux must be zero for resolving the linear equation~(\ref{eq:lineqn}). Numerical errors in solving the Poisson equation on the other hand will also introduce finite total magnetic flux as well. The calculated relative magnetic helicity using the previous method is -14441.45 and the value is -13253.36 using the new method. There is 8\% difference for the two methods.

\begin{table}[t]
\begin{center}
\caption{Results of the new scheme applied to the analytical model of \citet{low90} }\label{table:check}
\begin{tabular}{cccc}
\hline
\hline
side boundary& $\Phi_0^{\mathrm{a}}$ & $\oint{\bf A}_{\rm p} \cdot \rm d{\bf l}^{\mathrm{b}}$  & $\Phi^{\mathrm{c}}$\\
\hline
$z=0$   &  -3615.81  &  -3615.92   & -3508.19 \\
$z=l_z$ &   1461.13  &   1460.92   &  1490.70 \\
$x=0$   &  -1471.95  &  -1472.53   & -1575.36 \\
$x=l_x$ &  -1471.95  &  -1472.77   & -1576.94 \\
$y=0$   &   4006.42  &   4007.20   &  4091.27 \\
$y=l_y$ &   1068.27  &   1093.10   &  1070.95 \\

\hline
Total flux & -23.9037 &0.000610352 &     -7.57971     \\
\hline
\end{tabular}
\end{center}
\begin{list}{}{}
\item[$^{\mathrm{a}}$]{Magnetic flux through the side
boundaries of analytical model.}
\item[$^{\mathrm{b}}$]{The integral of
${\bf A}_{\rm p} \cdot \rm d{\bf l}$ taken along the edges of each side boundary.}
\item[$^{\mathrm{c}}$]{Magnetic flux obtained by the solution described in Sec. 2 for each side boundary.}
\end{list}
\end{table}

\begin{figure}[htb]
\centering
\includegraphics[angle=0,scale=0.5]{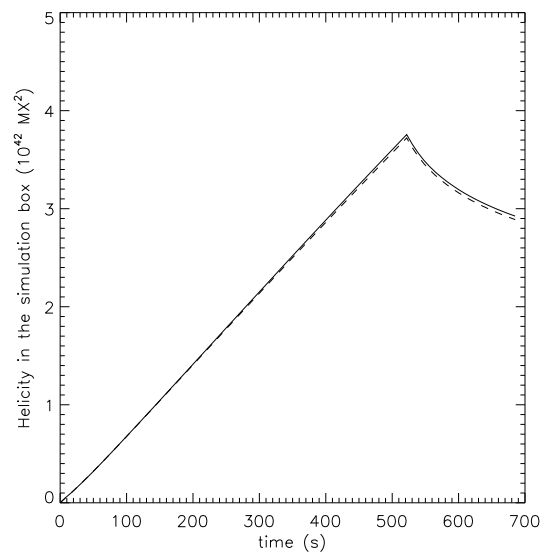}
\caption{Comparison of the magnetic helicity calculation by applying the previous and new scheme to the magnetic field data of ~\citet{Yang13}. The solid (dashed) line represents the new (previous) relative magnetic helicity evolution.}
\label{fig:RMHFigOld}
\end{figure}

We then applied our new scheme to re-calculate the relative magnetic helicity above active
region NOAA AR8210 in \citep{Yang13}. Fig.~\ref{fig:RMHFigOld} depicts the comparison of relative magnetic helicity evolution by using the previous and the new scheme.  The solid (dashed) line represents the new (previous) relative magnetic helicity evolution.  The
difference is very small (<1.2\%). This is due to the magnetic field structure of the corona above AR8210 obtained by \citep{Yang13} in which the fluxes through the side boundaries were almost balanced.

Finally, we apply our new scheme to the 3D magnetic field data obtained from simulated evolution of the solar corona above active region NOAA AR11429. The new GOEMHD3 code was used to reveal the magnetic field evolution in the solar atmosphere in response to the energy influx from the chromosphere through the transition region. The weak Joule current dissipation and  a finite viscosity is taken into account in the almost dissipationless solar corona. The GOEMHD3 code is a massively parallel code solving a second-order-accurate MHD equations \citep{Skala15}. It was successfully tested and applied to study the magnetic coupling between
the solar photosphere and corona based on multi-wavelength observations.
GOEMHD3 discretizes the ideal part of the MHD equations using a fast and efficient leap-frog scheme that is second-order accurate in space and time and whose initial and boundary conditions can easily be modified. For the investigation of diffusive and dissipative processes the corresponding terms are discretized by a DuFort-Frankel scheme. To always fulfill the Courant-Friedrichs-Lewy stability criterion, the time step of the code is adapted dynamically. Non-equidistant grids enhance the spatial resolution near
the transition region.
GOEMHD3 is parallelized based on a hybrid MPI-OpenMP programing paradigm, adopting a standard two-dimensional domain-decomposition approach. This allows investigate the long time evolution of the relative magnetic helicity in the solar Corona both in ideal and non-ideal magnetohydrodynamics by the non-ideal magnetohydrodynamical phase triggered by the switching on of resistivity.

\begin{figure*}[htb]
\centering
\includegraphics[angle=0,scale=1.0]{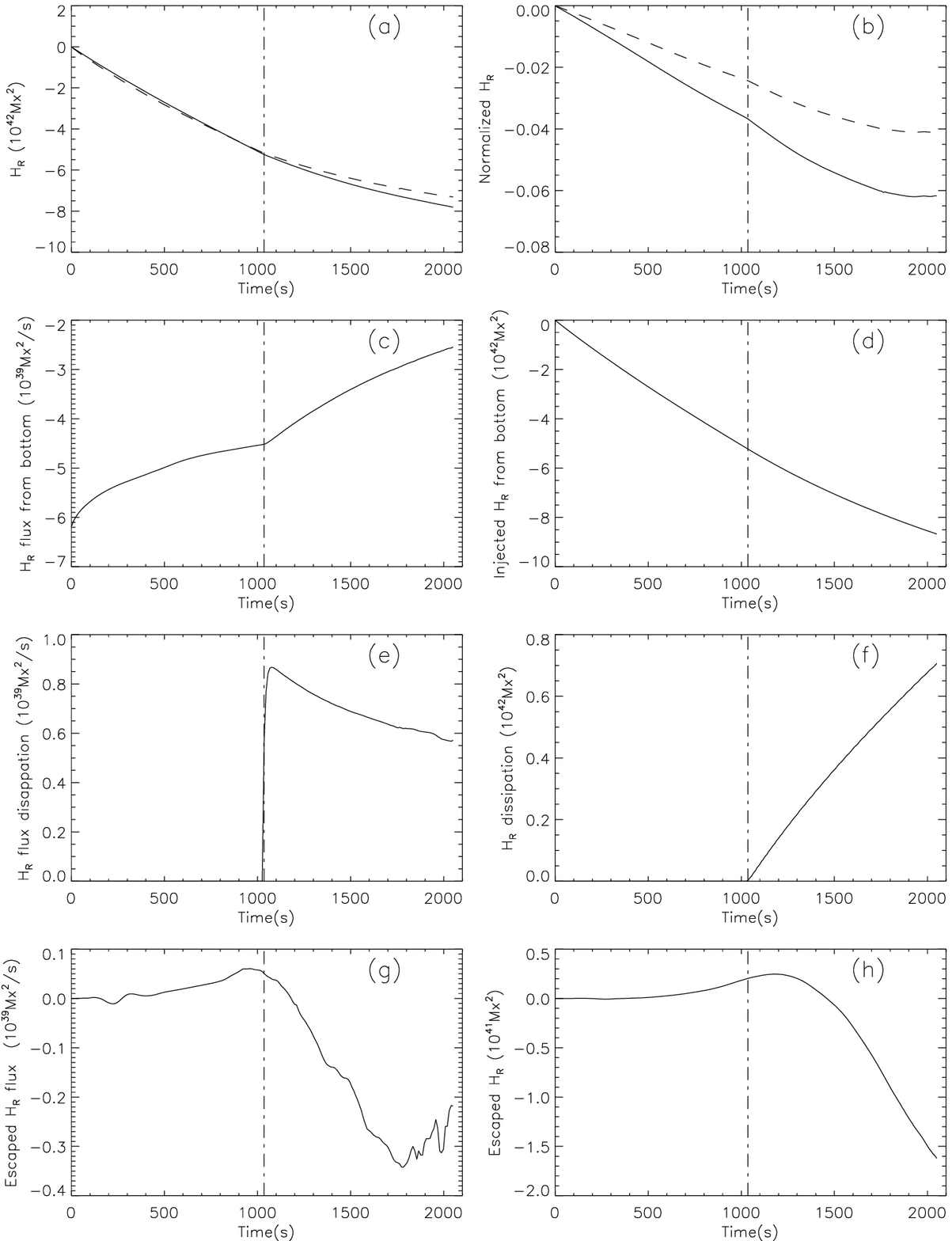}
\caption{Relative magnetic helicity evolution calculation using the result of GOEMHD3 Simulation. The dash-dot line represents the time when the fast reconnection started (t=1035s).
(a) Relative magnetic helicity in the simulation box (dashed line) and the accumulated relative helicity (solid line) sum of the helicity injection through the boundary and the dissipated in the volume helicity according to Eq.(6) in \citet{Yang13}.
(b) Evolution of the normalized relative helicity, in which the magnetic helicity is normalized by unsigned flux at the bottom boundary (solid line) and the six side boundaries (dashed line).
(c) Injected magnetic helicity flux through the bottom boundary.
(d) Injected magnetic helicity from the bottom boundary.
(e) Magnetic helicity dissipation rate in the simulation box.
(f) Magnetic helicity dissipation in the simulation box.
(g) Magnetic helicity change rate escaped from the other five boundaries except the bottom boundary.
(h) Magnetic helicity escaped from the other five boundaries except the bottom boundary.}
\label{fig:RMHFig}
\end{figure*}

Fig.\ref{fig:RMHFig} shows the evolution of relative magnetic helicity by using our
method to the simulation data from the application of GOEMHD3 to the evolution of the AR 11429 on March 07. 2012 \citep{Skala15}. The vertical  dash-dot line in the Fig.\ref{fig:RMHFig} indicate the time (t=1035s) when fast reconnection started. Fig.\ref{fig:RMHFig}a depicts the evolution of relative magnetic helicity in the simulation box (solid line) and the accumulated relative helicity (dashed line) calculated using the helicity change rate  Eq. (6) in the Yang et al. (2013).  The surface integral in Eq. (6) of \cite{Yang13} is extended to all six boundaries and the dissipative volume integral in Eq. (6) of \cite{Yang13} is also included in the computation.
Fig.\ref{fig:RMHFig}b depicts the normalized magnetic helicity evolution ($H_{\rm}/\Phi^2$). We use two types of fluxes $\Phi$ as the normalized parameter. One is unsigned magnetic flux only through the bottom boundary (solid line in Fig.~\ref{fig:RMHFig}b), which is usually used in observations. The other is unsigned magnetic flux through the six side boundaries (dashed line in Fig.~\ref{fig:RMHFig}b). A value of -0.038 (-0.025)  is reached after fast reconnection started while choosing unsigned magnetic flux through the bottom boundary (side boundaries) as the normalized parameter. The calculated normalized value -0.039 (-0.026) is consistent with the above simulation results if we substitute the two types of fluxes into the formula between the  magnetic helicity and the magnetic flux from the observations of newly emerging active regions \citep{Yang09b}:
\begin{equation}
\log\frac{\Delta H}{{H_0 }}=a\log \frac{{\Phi _m }} {{\Phi_0}}+b,
\label{eq:fluxHm}
\end{equation}
where $a=1.85$, $b=-0.41$, $H_0 = 10^{41}{\rm Mx}^2$, and $\Phi_0 =
10^{21}{\rm Mx}$.
, which describes the relation between the accumulated helicity $\Delta H$
and magnetic flux $\Phi _m$ for active regions. Fig.\ref{fig:RMHFig}c depicts the magnetic helicity flux through the bottom boundary into the solar corona. The total magnetic helicity flux is shown in Fig.\ref{fig:RMHFig}d. It is found that negative magnetic helicity is injected into the Corona due to the sub-photospheric plasma motion. The total injected helicity is
$-8.5\times10^{42}Mx^2$. Fig.\ref{fig:RMHFig}e depicts the  dissipation of magnetic helicity flux calculated as $-2\int{{\bf{E}}\cdot{\bf{B}}}dV$. The helicity dissipation is enhanced due to fast magnetic reconnection, i.e. non-ideal magnetohydrodynamics. The integrated dissipated magnetic helicity flux is shown in
Fig.\ref{fig:RMHFig}f. It is found that the sign of dissipation is positive, i.e. opposite to the helicity injected through the bottom boundary. The total magnetic helicity dissipation in the reconnection is $0.71\times10^{42}Mx^2$. Note that only
8\% of the injected magnetic is dissipated. This reflects the fact that
magnetic helicity is approximately conserved during fast magnetic reconnection as pointed
by \citep{Tay74}.
Fig.\ref{fig:RMHFig}g depicts the magnetic helicity flux which escape through the upper five boundaries. The total escaped magnetic helicity evolution is shown in
Fig.\ref{fig:RMHFig}h. It is found that only $-0.16\times10^{42} Mx^2$ of magnetic helicity flux escapes through the upper boundaries. This is
23\% of the dissipated helicity flux ($0.71\times10^{42}Mx^2$, see
Fig.\ref{fig:RMHFig}f) and only 2\% of the total injected magnetic helicity ( $-8.5\times10^{42}Mx^2$, see Fig.\ref{fig:RMHFig}d).  Note that only the helicity flux through all six boundaries is gauge invariant, and that fluxes though individual boundaries have values that may change by changing gauge \citep{par15}. In the calculation of Fig.\ref{fig:RMHFig} c-g at the separate boundary, we use the Coulomb gauge.

\section{Summary and Discussion}
\label{sec:summary}
We propose a new generalized scheme to calculate the vector
potential at the boundaries of a closed volume without the restrictions of the method applied in the previous paper \citep{Yang13}. We verified the new method using a analytical theoretical force-free model magnetic field  \citep{low90} and to the simulated data in \citet{Yang13}. We also apply the new method to simulated coronal
magnetic fields using the newly developed GOEMHD3 to investigate the magnetic helicity evolution of solar corona in the course of evolution of AR11429.

We found that only 8\% of the accumulated injected magnetic helicity is dissipated. This is consistent with the Taylor conjecture and also with the recent simulation results of \citet{par15}. Only 2\% of magnetic helicity injected through the bottom boundary escapes to the solar wind. This shows that the magnetic helicity cannot efficiently escape.
This may help to understand the lack of magnetic clouds \footnote{Near-Earth Interplanetary Coronal Mass Ejections Since January 1996: www.srl.caltech.edu/ACE/ASC/DATA/level3/icmetable2.htm. It is described in \cite{RC03,RC10}.} with considerable magnetic helicity in the interplanetary space \citep[Ref.][]{Dem15}. This is true even though several halo CMEs originated from the simulated active region AR11492.

Our simulation results further confirm that the absolute normalized magnetic helicity ($H_{\rm}/\Phi^2$) reaches 0.038 after the reconnection starts. It is interesting to note that recently an isolated and quickly decaying active region (NOAA 9729) was studied in detail as it passed across the solar disk. There was only one CME associated with that active region. This provided a good opportunity to investigate the consequences of single CME after the injection of magnetic helicity to the solar corona. The absolute value of normalized magnetic helicity was 0.036 just after solar eruption happened \citep{Yang15}. This is also similar to the obtained simulation results. A normalized helicity is one order of magnitude smaller than the value -0.16 (-0.18) obtained by the MHD simulation results of kink or torus instability of ideal MHD \citep{Fan07} when eruption happened. This might be due to the anomalous resistivity caused by the strong micro-turbulence and microscopic structures \citep{bue06b} used in the GOEMHD3
model. The anomalous current dissipations allows essentially increase the dissipation rate of magnetic energy following a global MHD instability (B\"uechner et al. 2017, in preparation).

\cite{Zhang06} as well as \cite{Zhang08} proposed that there is an upper bound of the total magnetic helicity that a force-free field can contain in a multipolar force-free magnetic field structure before eruption, which is very close to the obtained value above. However, the allowable level of helicity in a force-free field is only a sufficient condition for an eruption. A CME expulsion may still occur even before this helicity limit is reached \citep{Zhang06}. The upper-bound normalized magnetic helicity value deviates evidently for different magnetic structures in the theoretical force-free field model. For multipolar fields, the helicity upper bound
can be 10 times smaller than that of a dipolar field \citep{Zhang08}.  But in our observations and simulation result, the normalized helicity reaches the same order when eruption happens. It is essential to investigate the evolution of magnetic helicity and energy using more observations and simulations in the future study.

Above all, after introducing a new scheme removing the former restriction on the magnetic flux through the boundaries, we can calculate the relative magnetic helicity of any magnetic field structure in Cartesian coordinates. In the observations, we could use a force-free
extrapolation to obtain the three-dimensional magnetic structure to analyze the evolution of relative magnetic helicity.

\begin{acknowledgements}
We would like to thank the referee for carefully reading our
manuscript and for giving such constructive comments which
substantially helped improving the paper. This study is supported by grants 10733020, 10921303,
41174153,11173033 11178016 and 11573037 of National Natural Science
Foundation of China, 2011CB811400 of National Basic Research Program
of China, a sandwich-PhD grant of the Max-Planck Society and the
Max-Planck Society Interinstitutional Research Initiative Turbulent
transport and ion heating, reconnection and electron acceleration in
solar and fusion plasmas of Project No. MIF-IF-A-AERO8047.The
authors also like to thank the Supercomputing Center of Chinese
Academy of Sciences (SCCAS) and Max Planck Computing and Data Facility (MPCDF)
for the allocation of computing time.
\end{acknowledgements}

\end{document}